\documentclass[lettersize,journal]{IEEEtran}
\usepackage{amsmath,amsfonts}
\usepackage{algorithmic}
\usepackage{algorithm}
\usepackage{array}
\usepackage[caption=false,font=normalsize,labelfont=sf,textfont=sf]{subfig}
\usepackage{textcomp}
\usepackage{stfloats}
\usepackage{url}
\usepackage{verbatim}
\usepackage{graphicx}
\usepackage{booktabs}
\usepackage{cite}
\usepackage{xspace} 
\begin{document}

\title{An Immersive Virtual Reality Bimanual Telerobotic System With Haptic Feedback}

\author{Han Xu$^{1}$, Mingqi Chen$^{1}$, Gaofeng Li$^{2}$,  Lei Wei$^{3}$, Shichi Peng$^{1}$, Haoliang Xu$^{1}$, Qiang Li$^{1*}$

\thanks{$^{1}$Han Xu, Mingqi Chen, Shichi Peng, Haoliang Xu and Qiang Li are with Dexterous Robotics Lab, Shenzhen Technology University, Shenzhen, 518118, China}%
\thanks{$^{2}$Gaofeng Li is with the College of Control Science and Engineering, Zhejiang University, Hangzhou, 310027, China.   
        {\tt\small gaofeng.li@zju.edu.cn}}%
\thanks{$^{3}$Lei Wei is with the Institute for Intelligent Systems Research and Innovation (IISRI), Deakin University. 75 Pigdons Road, Waurn Ponds, VIC 3216, Australia.
        {\tt\small lei.wei@deakin.edu.au}}%
\thanks{*Corresponding author (liqiang1@sztu.edu.cn)}
   
}
\maketitle    
\begin{abstract}
 
In robotic bimanual teleoperation, multimodal sensory feedback plays a crucial role, providing operators with a more immersive operating experience, reducing cognitive burden, and improving operating efficiency. 
In this study, we develop an immersive bilateral isomorphic bimanual telerobotic system, which comprises dual arm and dual dexterous hands, with visual and haptic force feedback. To assess the performance of this system, we carried out a series of experiments and investigated the user's teleoperation experience. The results demonstrate that haptic force feedback enhances physical perception capabilities and complex task operating abilities. In addition, it compensates for visual perception deficiencies and reduces the operator's work burden. Consequently, our proposed system achieves more intuitive, realistic and immersive teleoperation, improves operating efficiency, and expands the complexity of tasks that robots can perform through teleoperation.

\end{abstract}
\section{Introduction}
\IEEEPARstart{I}{n} many application scenarios, bimanual robots are required to perform tasks on behalf of humans, such as search and rescue, underwater operations, and elderly care \cite{ref1,ref2,ref3,10572265}. Teleoperation of bimanual robots combines human intelligence and expertise with the robot's physical abilities \cite{10035484}. Leveraging human expertise and skills by teleoperation to control robots enables the transfer of human operational capabilities to remote robots, allowing for more complex and precise tasks to be performed.

To do so, sufficient sensory feedback is required by operator to perceive, understand and reason about the environment of the remote robot. Currently, a common approach is to utilize Virtual Reality(VR) head-mounted displays(HMD) and stereo cameras to provide operators with stereo visual feedback from the remote side, presenting spatial information in a 3D stereoscopic manner \cite{naceri2021vicarios, gallipoli2024virtual, smith2024augmented, cheng2024open, wei2021multi}. However, this is insufficient in many applications, and the sole reliance on visual feedback will lead to limitations in teleoperation, which also diminishes the human-robot interaction experience \cite{rea2022still, spano2021teleoperating}. 

\begin{figure}[tbp]  
    \centering        
    \includegraphics[width=0.48\textwidth]{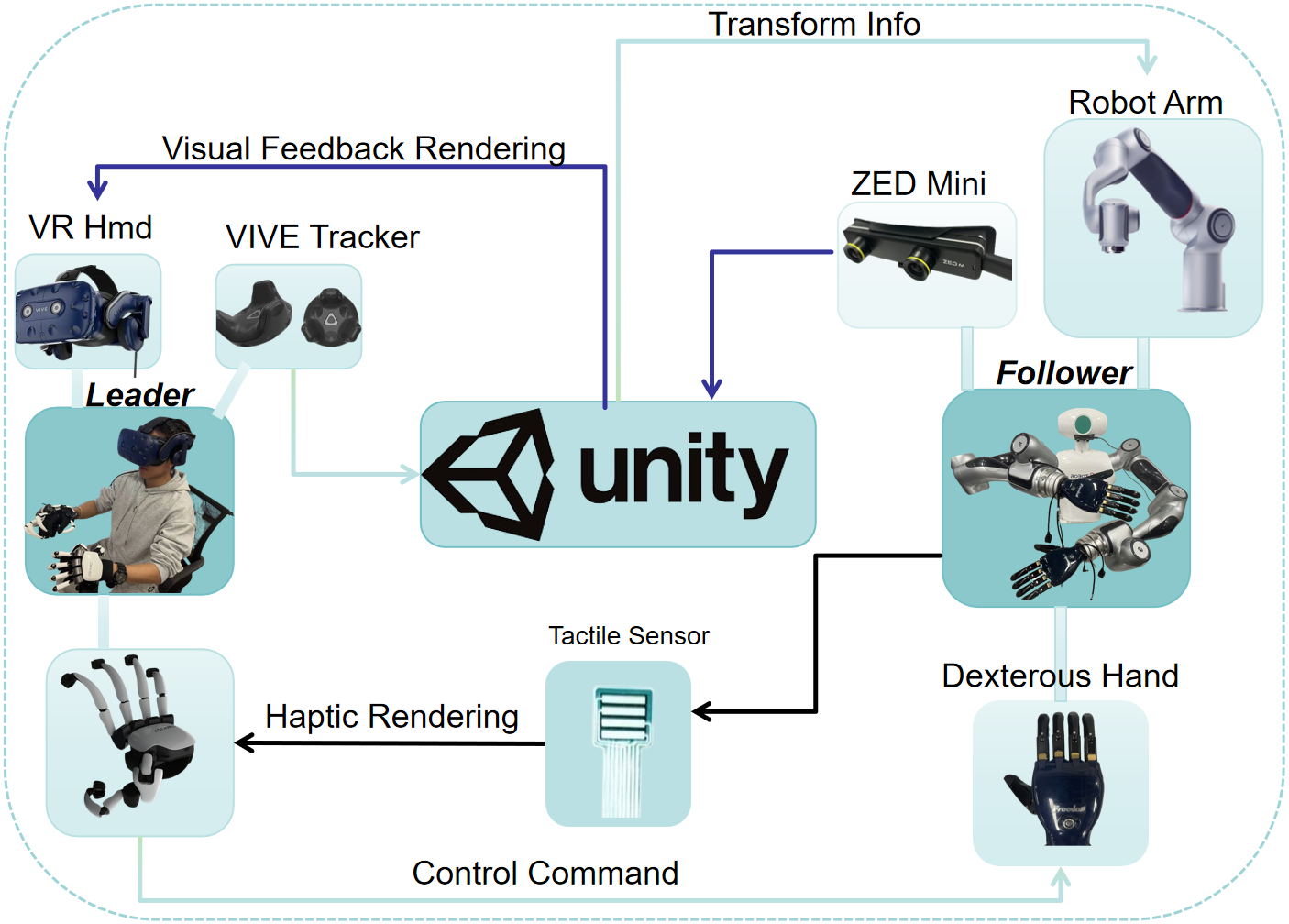}  
    \caption{The proposed system for immersive and intuitive teleoperation with bilateral haptic feedback, and visual rending, isomorphic dexterous arm and hand control.}  
    \label{fig:system}  
\end{figure}
As one of the five senses for perceiving properties and motion, haptic plays a vital role in human interaction with the surrounding world. While vision allows operators to acquire external features of objects, such as size, shape, pose, and surface smoothness\cite{wang2021tactual,ficuciello2019vision,du2021vision} , the internal attributes of objects (e.g., weight, softness, fragility) can only be perceived through haptic. By incorporating haptic modality, operators can better perceive the surrounding environment and interact with it, experiencing a real-time sensation from the remote robot's feedback at the physical level. In certain scenes, limited lighting, occlusion, or focus can make it challenging to acquire geometric properties and positions in the environment. The haptic can well compensate for this deficiencies of visual modality. In other case, haptic and visual modalities can work together to play a role, for example, high-quality grasping can be achieved by judging the geometric shape of soft objects through vision and the degree of deformation through haptic to determine a reasonable grasping position,.

In bimanual robot control, mapping the robot's haptic force feedback to the operator's skin in an isomorphic manner can further enhance immersion and interactivity. By designing intuitive human-robot interfaces that incorporate wearable haptics, we can achieve a more immersive and interactive teleoperation experience, enabling operators to feel more connected to the remote robot and its environment.
\section{Related work}
In the field of robot teleoperation, researchers have explored various methods to enhance human-robot interaction. For instance, Wei {\it et al.} \cite{wei2021multi} proposed a multi-view visual feedback method in teleoperation, which was shown to improve the teleoperation performance of single-arm robots. Zhu {\it et al.} \cite{zhu2023visual} estimated the force of a visual-haptic sensor in teleoperation, enabling complex hand-in-hand spinning using a 2 degree-of-freedom(DoF) Omega gripper. Fan {\it et al.} \cite{fan2023digital} developed a Mixed Reality(MR) interface for single-arm gripper teleoperation. However, the use of single-arm and gripper-based telerobotic systems have limitations, as they provide a less natural and intuitive way of human-robot interaction, and are often restricted to low-degree-of-freedom scenarios.

Cheng {\it et al.} \cite{cheng2024open} proposed OpenTeleVision, an immersive telerobotic system that enables remote control of a humanoid robot by mirroring the operator's arm and hand movements, and provides training data for imitation learning. Fu{\it et al.} \cite{fu2024humanplus} used motion capture to map human movements to a humanoid robot, allowing more intuitive control. Aliyah Smith {\it et al.} \cite{smith2024augmented} designed an augmented reality interface using Microsoft HoloLens device, which aimed to reduce the operator's task burden. Although these studies have incorporated visual feedback to perceive visual information of the remote environment, relying solely on visual feedback has its limitations. Moreover, single-modal sensory feedback restricts the acquisition of remote information, leading to a less immersive teleoperation experience. In contrast, multi-modal sensory feedback has the potential to increase the immersion of teleoperation compared to single-modal sensory feedback.

There are several general methods for providing haptic force feedback using vibration actuators. For example, Liu {\it et al.} \cite{8794230} employed a network of vibration motors to deliver haptic feedback in VR grasping applications. Similarly, Ding {\it et al.} \cite{Ding2024BunnyVisionProRB} designed a haptic interface that utilizes vibrations to simulate the haptic sensations experienced by a robot in a remote environment. Although these approaches offer detailed rendering in a specific way, the use of vibration provides only a pseudo-force feedback when interacting with the environment. In contrast, our goal is to introduce grounded haptic feedback in teleoperation.

Given the downside of current available teleoperation, we propose a new bimanual telerobotic system and the contributions are summarized as following:

\begin{enumerate}

\item{Our approach combines haptic and visual rendering in bimanual robot teleoperation, providing a more immersive experience by enabling multimodal perception of remote visual and touching information.}
\item{We assess the efficacy of haptic feedback in enhancing the performance of bimanual telerobotic system, focusing on aspects such as blind grasping, haptic sensation of operator, in-hand manipulation, and manipulation efficiency.}
\end{enumerate}
\section{Telerobotic System}
\subsection{System Setup}
Our proposed teleoperation framework, as illustrated in Fig.\ref{fig:system}, integrates a human-robot interface with bilateral haptic feedback, a visual interface for 3D rendering of the remote scene, and a control module for tele-operating robotic arm and dexterous hand. 
In the proposed system, the bimanual robot, located at the remote end, comprises 2 DoF robotic arms and two 6 DoF five-fingered dexterous hand\textsuperscript{1} with tactile sensor on each finger tip. A ZED mini\textsuperscript{2} camera is mounted on the robot's head, providing 3D visual information of the remote environment.
On the operator side, we utilize an HTC VIVE PRO\textsuperscript{3} VR device, which consists of a HMD and two VIVE Trackers\textsuperscript{4} for tracking the operator's wrist pose. The HMD provides stereoscopic images to the operator, rendering the remote scene in 3D. The operator wears a dexmo exoskeleton glove\textsuperscript{5}, which has a force feedback motor on each finger, with a maximum torque of 0.5 N·m, to provide haptic feedback. 
\footnotetext[1]{www.tsingrbytech.com}
\footnotetext[2]{https://www.stereolabs.com/en-jp/store/products/zed-mini}
\footnotetext[3]{https://www.VIVE.com/cn/product/VIVE-pro/}
\footnotetext[4]{https://www.VIVE.com/cn/accessory/tracker3/}
\footnotetext[5]{https://www.dextarobotics.com/}

\begin{figure*}[t] 
\centering 
\includegraphics[scale=0.58]{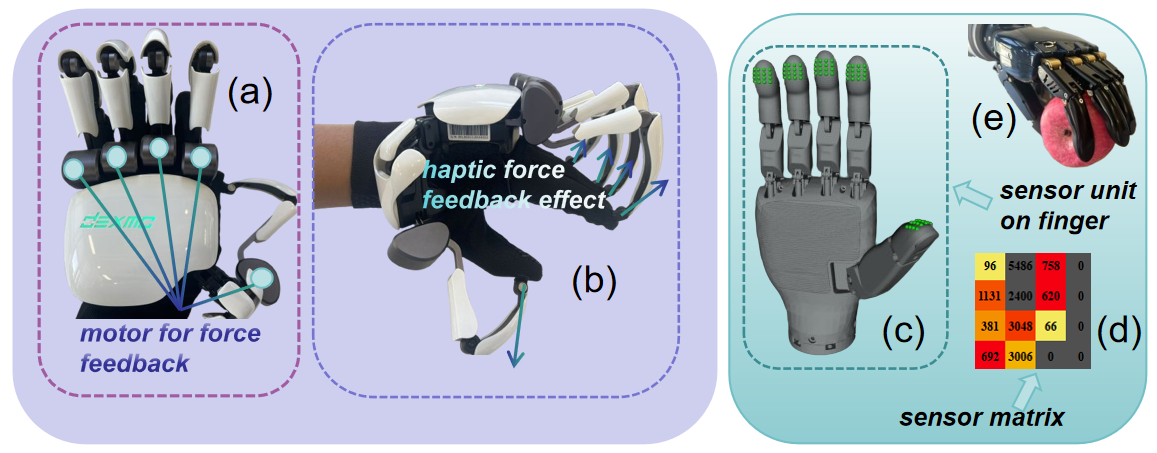} 
\caption{Illustration of haptic rendering. (a) Each finger of the Dexmo exoskeleton glove is equipped with a motor to provide haptic feedback. (b) The human operator senses haptic force feedback through the Dexmo exoskeleton glove. (c) A tactile sensor with 16 sense units is built into the dexterous hand. (d) Example of a haptic sensor matrix when a sensor is triggered. (e) Grasping example.} \label{fig:hapticinterface} \end{figure*}
\subsection{Haptic Interface}
\begin{figure}[htbp]  
    \centering        
    \includegraphics[width=0.48\textwidth]{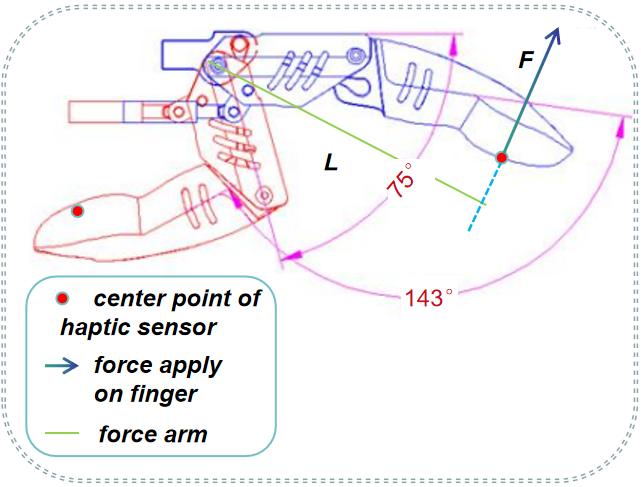}  
    \caption{Illustration of the torque calculation. The $F$ refers the force applying on fingertip (see \eqref{eq:F1}). The $L$ is the force arm.}  
    \label{fig:torque}  
\end{figure}

As show in Fig. \ref{fig:hapticinterface}. The robotic hand is equipped with a tactile array at each fingertip to perceive tactile information. Each sensor array, comprising 4$\times$4 taxels, displays the magnitude of the received force (in 1/3000 N increments). We represent the taxel readout as $M_{ij}$, where the force $F$ is the sum of the force matrix:
\begin{equation}
    \label{eq:F}
    F = \sum_{i=1}^{4} \sum_{j=1}^{4} M_{ij}
\end{equation}
The motors in the wearable exoskeleton gloves render the haptic force feedback for the operator. Assuming that point of action is at the center point of the tactile sensor and the vertical distance from this point to the proximal phalanx of the finger is set as the force arm $L$, as Fig. \ref{fig:torque}, the corresponding torque of the finger $\tau$ can be calculated as:
\begin{equation}
    \label{eq:tor}
\tau = F \cdot L
\end{equation}

In haptic interface, the robotic hand perceive and send torque information, and the wearable exoskeleton glove receive the information. These two are connected through Transmission Control Protocol (TCP) communication. This, synchronously, creates an intuitive haptic sensation within the teleoperation scenario, which enable haptic telepresence sensation. Moreover, to ensure compatibility with the robotic hand's tactile sensor limited communication performance, our haptic feedback module operates at a frequency of 62Hz.
\subsection{Visual Interface}
For the operator side, we utilize a commercially available VR system HTC VIVE PRO to provide 3D immersive visual rendering of remote global scene. The HMD has two displays which are set in a stereo manner. Visual 3D effect rendering can be created by presenting two images with parallax to each display. 
For the robot side, we use ZED mini camera to acquire visual information of the remote scene. The ZED mini is a stereo camera, where the distance between its left and right cameras closely matching the inter-pupillary distance of a human. 

Visual rendering is done by open-source libraries. They mainly include \textit{pyzed.sl}\textsuperscript{6} and \textit{Zed Unity Plugin v3.8.0}\textsuperscript{7}. Using the \textit{pyzed.sl} library, stereo video stream from the ZED Mini camera can be captured in a remote scene and sent via TCP communication. Importing the \textit{Zed Unity Plugin v3.8.0 package} into Unity and adding \textit{ZED manager}, Unity is able to receive the video stream. In this way, the remote video stream can be transmitted from the robot side to the operator side in the LAN.

We use Unity game engine for scene rendering. With this engine we are able to provide a high quality immersive experience to the user through HTC VIVE HMD. First, we create a \textit{camera} Object in Unity that corresponds to the HMD. Next, we create two \textit{canvases} in the \textit{camera} and then render the video stream which contain parallax from two cameras of ZED mini to two \textit{canvases} in real-time. Last we map and broadcast the canvas to the corresponding display of the HTC VIVE headset. When the operator wears HMD, they can perceive 3D environment surrounding the robot, enable visual telepresence sensation.

\footnotetext[6]{https://github.com/stereolabs/zed-python-api}
\footnotetext[7]{https://github.com/stereolabs/zed-unity/releases/}
\subsection{Dual Arm and Dexterous Hand Control}

\begin{figure}[tbp]
  \centering
  \includegraphics[width=0.48\textwidth]{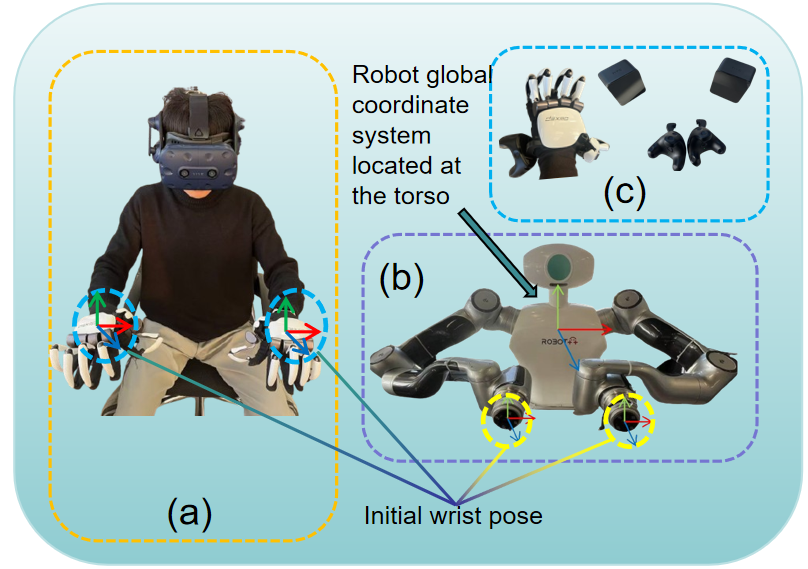}
  \caption{(a) and (b), the position and orientation of the bimanual robot's arm end-effector are controlled through the relative position and orientation of operator's wrist. (c) The dexmo exoskeleton glove and VIVE Tracker are bound together, and the wrist position and orientation are read through the locator. The joint angles of the 7-DoF robotic arm are calculated using traditional inverse kinematics algorithms.}
  \label{fig:trans}
\end{figure}

The relevant coordinate system of the arm and hand control are illustrated in Fig. \ref{fig:trans}. We control the robotic arm by calculating the operator's wrist pose relative to the initial pose. By binding the VIVE Tracker and exoskeleton glove (Fig. \ref{fig:trans}(c)), we can obtain the position and orientation of the operator's wrist. We capture the initial position and orientation of the operator's wrist relative to the Unity global coordinate system as$^{G}\mathbf{P}_{init}$ and $_{L}^{G}\mathbf{R}$ (Fig. \ref{fig:trans}(a)). The initial pose of the operator's wrist is employed as a local reference coordinate system to calculate the relative motion of the wrist with respect to its initial pose. Following, we capture the real-time position and orientation of the operator's wrist relative to the Unity global coordinate system as $^{G}\mathbf{P}_{now}$ and $_{N}^{G}\mathbf{R}$. The position $^{L}\mathbf{P}$ of the operator's wrist in the local coordinate system is:

\begin{equation} 
\label{eq:a} 
^{L}\mathbf{P} =  _{L}^{G}\mathbf{R}^{-1} \cdot (^{G}\mathbf{P}_{now} - ^{G}\mathbf{P}_{init})
\end{equation}

The orientation $_{N}^{L}\mathbf{R}$ of the operator's wrist in the local reference system is given by:

\begin{equation} 
\label{eq:b} 
_{N}^{L}\mathbf{R} =  [^{L}\mathbf{X}_{N}, ^{L}\mathbf{Y}_{N}, ^{L}\mathbf{Z}_{N}]
\end{equation}
Where $^{L}\mathbf{X}_{N}$, $^{L}\mathbf{Y}_{N}$ and $^{L}\mathbf{Z}_{N}$ represent the unit vectors of the wrist's coordinate system along the principal axes in local coordinate system. These unit vectors are obtained through the following transformations:
\begin{equation} 
\label{eq:c} 
^{L}\mathbf{X}_{N} = ^{G}_{L}\mathbf{R}^{-1} \cdot ^{G}\mathbf{X}_{N}
\end{equation}
\begin{equation} 
\label{eq:d} 
^{L}\mathbf{Y}_{N} = ^{G}_{L}\mathbf{R}^{-1} \cdot ^{G}\mathbf{Y}_{N}
\end{equation}
\begin{equation} 
\label{eq:e} 
^{L}\mathbf{Z}_{N} = ^{G}_{L}\mathbf{R}^{-1} \cdot ^{G}\mathbf{Z}_{N}
\end{equation}
Where $^{G}\mathbf{X}_{N}$, $^{G}\mathbf{Y}_{N}$ and $^{G}\mathbf{Z}_{N}$ represent the unit vectors of the wrist's coordinate system along the principal axes in Unity global coordinate system. Consequently, we get the position $^{L}\mathbf{P}$ and orientation $^{L}_N\mathbf{R}$ of the operator's wrist with respect to the initial pose, and transfer it as control signal to the robot side.

As show in (Fig. \ref{fig:trans}(b)), the global coordinate system on the robot side is located at the robot's torso. Before the teleoperation command is issued, the robotic arm's end-effector is pre-positioned at position $^{G}\mathbf{K}_{init}$ and orientation $^{G}_{L}\mathbf{Q}$ relative to the robot global coordinate system. When the control signal is received by the robot side, the expected position $^{G}\mathbf{K}_{now}$ and orientation $^{G}_{N}\mathbf{Q}$ of the robot's wrist end-effector relative to the global coordinate system are:
\begin{equation} 
\label{eq:f} 
^{G}\mathbf{K}_{now} = ^{G}_{L}\mathbf{Q} \cdot   ^{L}\mathbf{P} + ^{G}\mathbf{K}_{init}
\end{equation}
\begin{equation} 
\label{eq:g} 
^{G}_{N}\mathbf{Q} = [^{G}\mathbf{A}_{N}, ^{G}\mathbf{B}_{N}, ^{G}\mathbf{C}_{N}]
\end{equation}
\begin{equation} 
\label{eq:h} 
^{G}\mathbf{A}_{N} = ^{G}_{L}\mathbf{Q} \cdot ^{L}\mathbf{X}_{N}
\end{equation}
\begin{equation} 
\label{eq:i} 
^{G}\mathbf{B}_{N} = ^{G}_{L}\mathbf{Q} \cdot ^{L}\mathbf{Y}_{N}
\end{equation}
\begin{equation} 
\label{eq:j} 
^{G}\mathbf{C}_{N} = ^{G}_{L}\mathbf{Q} \cdot ^{L}\mathbf{Z}_{N}
\end{equation}
Where $^{G}\mathbf{A}_{N}$, $^{G}\mathbf{B}_{N}$ and $^{G}\mathbf{C}_{N}$ represent the unit vectors of the robot wrist's coordinate system along the principal axes, with respect to the robot torso global coordinate system. 
As a result, the pose [$^{G}_{N}\mathbf{Q}  $,$^{G}\mathbf{K}_{now}$] of the robot's end-effector can be defined based on the relative position of the operator's wrist according to the above equation.

\begin{figure}[tbp]
  \centering
  \includegraphics[width=0.48\textwidth]{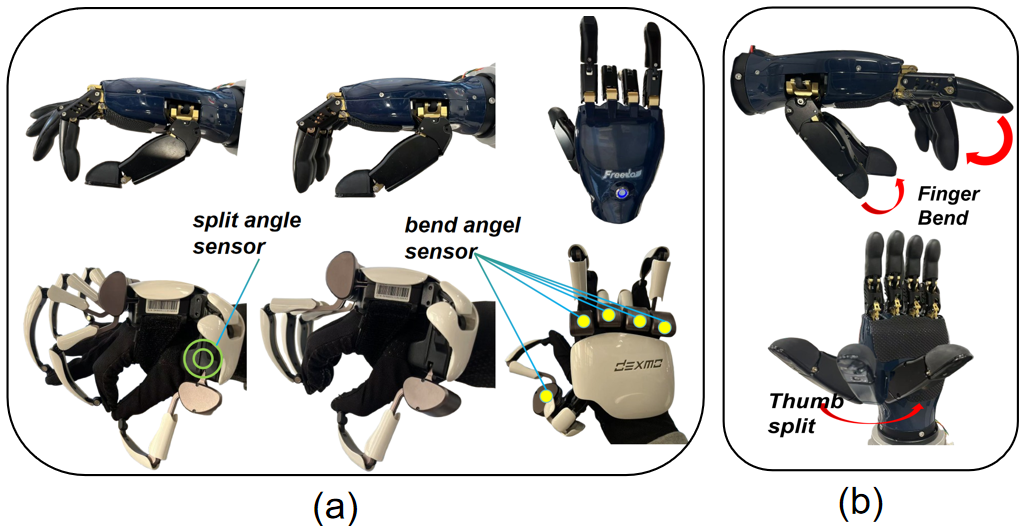}
  \caption{(a) the bend and split angles are read by sensors built in exoskeleton glove. The info obtained by sensors will be used to control the dexterous hand for isomorphic teleoperation. (b) dynamic examples of finger bending and thumb splitting.}
  \label{fig:hm}
\end{figure}

For hand control, we adopt isomorphic mapping. Each finger of the wearable exoskeleton glove is equipped with sensors to measure its bending angle, while the split angle of thumb is also measured. The wearable gloves connects to the host computer via wireless connection, and the host computer transmits the data to the robot side via TCP communication. The dexterous hand's finger bending and thumb split are controlled to match the values read by the exoskeleton glove sensors. Through this approach, the operator can achieve simple and efficient control of the robot, while adapting to the force feedback to adjust hand control. The control effect is shown in Fig. \ref{fig:hm}(a) and the dynamic examples of finger bending and thumb splitting is shown in Fig. \ref{fig:hm}(b). The control frequency of the hand control module is 500Hz.
\section{Experiment And Result}
To evaluate the performance of the proposed system, we designed a series of experiments. In experiment A, we selected several objects with varying hardness levels and executed two dynamic operation tasks to demonstrate the haptic interface's capability to convey force feedback and perceive internal properties of objects. Experiment B, we executed a blind grasping task to demonstrate the ability of haptic feedback to compensate for the limitations of visual feedback under occlusion. Experiment C, we designed comparison experiments to evaluate the performance of teleoperation tasks with different feedback: single visual feedback and visual plus haptic feedback. These experiments aimed to demonstrate the improvement of teleoperation performance because the introducing of the haptic feedback. In experiment D, we conducted subject studies to investigate the impact of multimodal sensory feedback on teleoperation efficiency and immersion.
\subsection{Haptic Force Feedback Display}
\begin{figure}[tbp]  
    \centering        
    \includegraphics[width=0.48\textwidth]{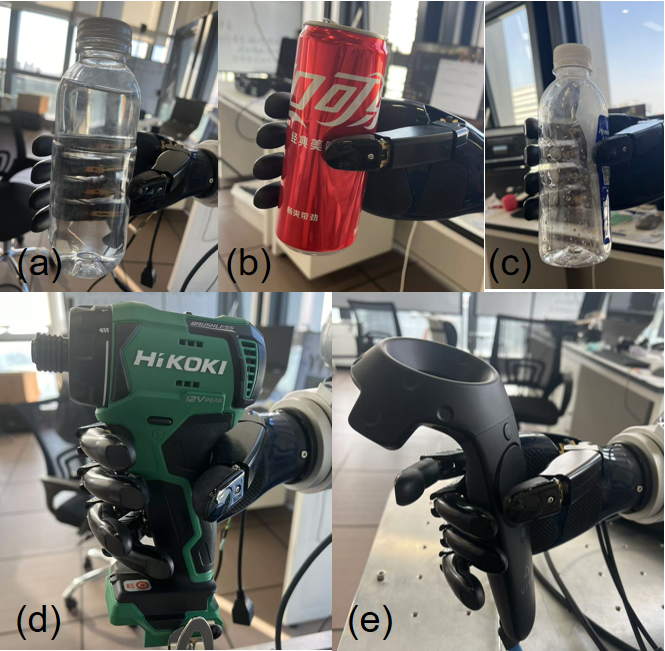}  
    \caption{(a) hard bottle grasping. (b) empty coke grasping. (c) soft bottle grasping. (d) electric drill dynamically operating. (e) VR controller dynamically operating.}  
    \label{fig:graspscene}  
\end{figure}
\begin{figure}[tbp]
  \centering
  \includegraphics[width=0.48\textwidth]{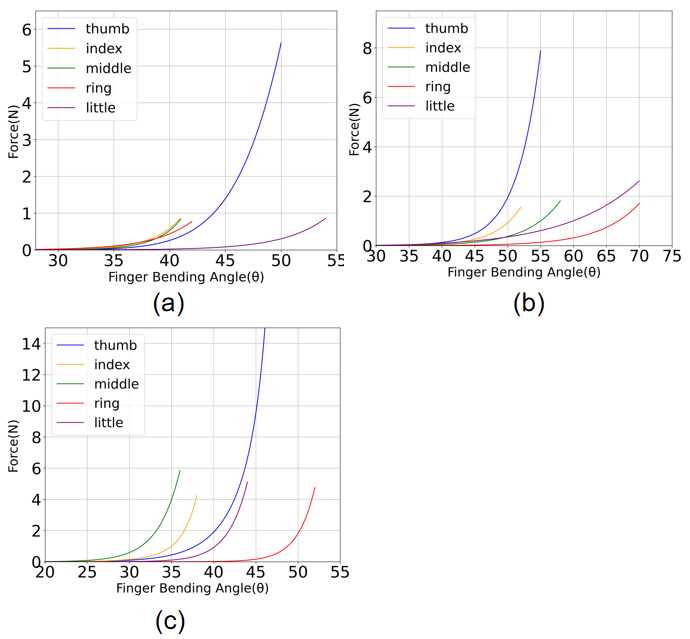}
  \caption{Visualization of operator's haptic experience. For different object, the curve of each finger has different slope, which reflect the object's hardness. {(a) empty coke bottle grasping. (b)soft bottle grasping. (c) hard bottle grasping. For objects with different hardness, the force-finger bending angle curves changes differently. The harder the object, the greater the force change with the finger bending angle, and the steeper the slope of the curve.}}
  \label{fig:forcedis}
\end{figure}

Grasping and dynamic operation tasks are performed to showcase its haptic feedback capabilities. For object grasping, we selected three bottles of varying hardness: an empty coke bottle Fig. \ref{fig:graspscene}(a), a soft bottle Fig. \ref{fig:graspscene}(b) and a hard bottle Fig. \ref{fig:graspscene}(c). During grasping, the bending angles of robotic hand's finger were changed in response to the control commands from the operator side. The force-displacement relationship in robotic hand side was different due to the varying hardness of the object, and this resulted that the changing of computed torque was different for the varying objects. When the torque is rendered in the exoskeleton glove, the operator could clearly perceive the hardness differences between the different bottles. It is also important to recognize that the hardness of an object is not a fixed property, as traditionally believed. Instead, the perceived hardness is a complex and variable attribute that is influenced by factors such as grasping position, finger contact, and other real-world conditions. 
To visualize the results, we recorded robotic tactile force and corresponding robotic finger bending angles. The force-finger bending angle curves exhibited significant differences among the objects in Fig.\ref{fig:forcedis}.
\begin{figure}[t]

\centering
\includegraphics[width=0.48\textwidth]{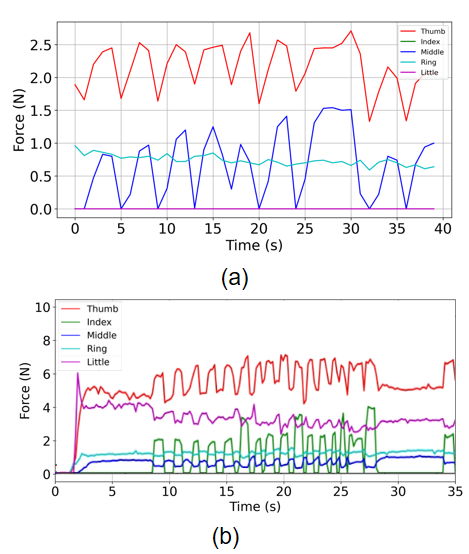}
\caption{Force variation with bending angle in dynamic operation. (a) VR controller dynamically operating. (b) Electric drill dynamically operating. Through the haptic feedback interface, the operator can sense the real-time dynamic changes of force feedback during the operating. All fingers contact forces are visualized.}
\label{fig:forcetime}
\end{figure}

For the dynamic operation task, we selected two tasks: drill operation Fig. \ref{fig:graspscene}(d) and VR controller operation Fig. \ref{fig:graspscene}(e). In the first task, the robotic hand was controlled to hold a VR controller and repeatedly pressed the control button. In the second task, the operator was asked to hold a drill and pressed the drill control key with the index finger. The tactile forces were recorded and shown in Fig. \ref{fig:forcetime} for different fingers to reflect the changes during the dynamic operation tasks. With the help of haptic feedback, the operator can clearly sense the change of the contact force of the remote robotic hand.
  
In VR controller operation, the operator use the ring finger to fix the object, and the force feedback of the ring finger is relatively stable. The thumb and middle finger are placed opposite to each other when the thumb press the controller. In electric drill operating, the ring finger and little finger mainly play a fixed role, and force feedback of the two fingers is relatively stable throughout the process. The thumb is placed opposite to the other four fingers. When the index finger pressed the button, force of the index finger increased. The thumb and middle also changed in order to adapt to the changes of the index finger. The coordination between the middle finger and the index finger is weaker, and the change in force of middle finger is also weaker.

\subsection{Blind Grasping Experiment}
Extensive research has shown that haptic feedback is useful for blind grasping. In teleoperation, when vision is limited, we will not be able to determine the accurate position and orientation of the object, which will lead to difficulties in operation. Haptic feedback can compensate these limitations and assist the robot in successfully executing these tasks.
\begin{figure}[htbp]
  \centering
  \includegraphics[width=0.48\textwidth]{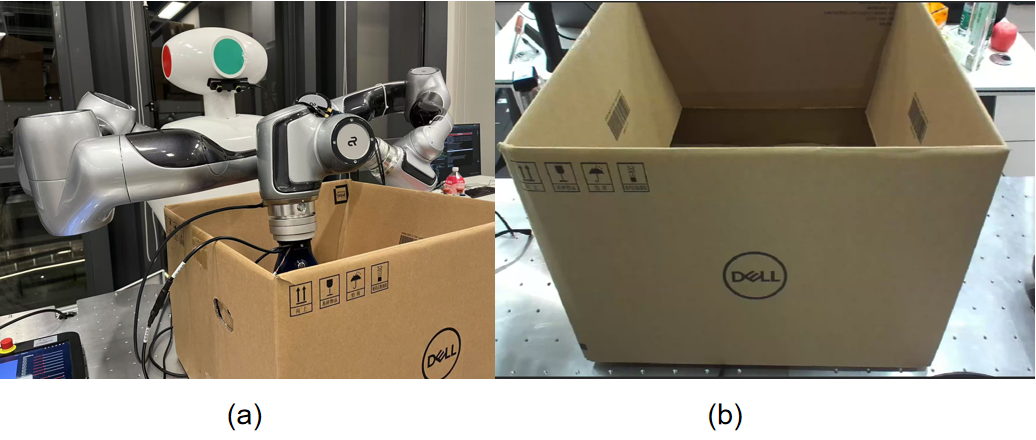}
  \caption{Blind grasping experiment. The object inside the box is completely occluded and invisible.}
  \label{fig:blind}
\end{figure}

To evaluate the effectiveness of haptic feedback in compensating visual information, we designed a blind grasping experiment. A box with an open top was placed in front of the ZED mini camera, and the contents of the box were invisible to the operator in Fig. \ref{fig:blind}. 

Without haptic feedback, the tele-grasping was not possible. With the haptic feedback, the operator can uses haptic feedback to sense the position of the object to complete the grasp. We randomly selected three office objects (bottle, punch machine and extension cord) and implemented the grasping operation for 25 times. The success rate are 0.56, 0.52 and 0.44 respectively. The results show that the haptic feedback compensated the limitations of visual feedback to some extent. We analyzed the grasping failure reason with the haptic feedback, and it is mainly caused by limitation of the sensitivity of the sensor and limited sensor distribution in the robotic hand.  
\subsection{Comparison Experiment} 
\subsubsection{Active Sliding Experiment}
\begin{figure}[tbp]
  \centering
  \includegraphics[width=0.48\textwidth]{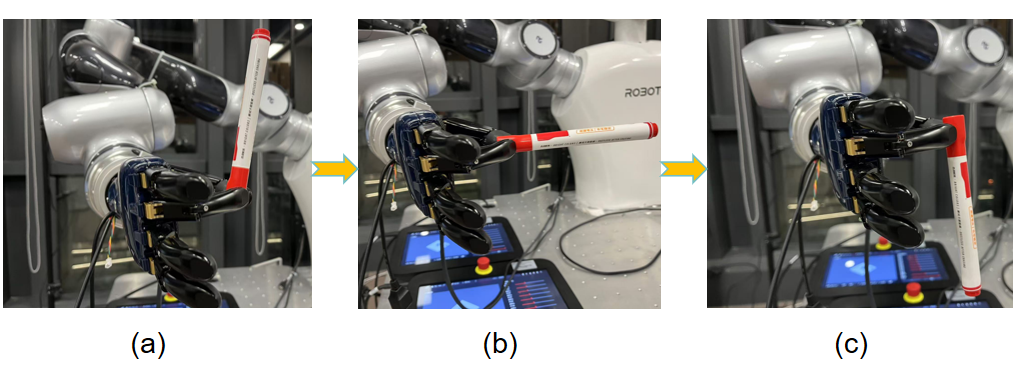}
  \caption{Active Sliding Experiment. (a) Initial state. (b) Intermediate state. (c) Final state.}
  \label{fig:exper}
\end{figure}
To assess the impact of haptic feedback on the performance in fine manipulation, we conducted an active sliding experiment. The robotic hand was teleoperated to control one object's in-fingers twist motion. The experiment show that haptic feedback compensated the human vision-based teleoperation well, and success rate of in-hand fine manipulation is improved largely.

\begin{figure}[t]
\centering
\includegraphics[scale=0.35]{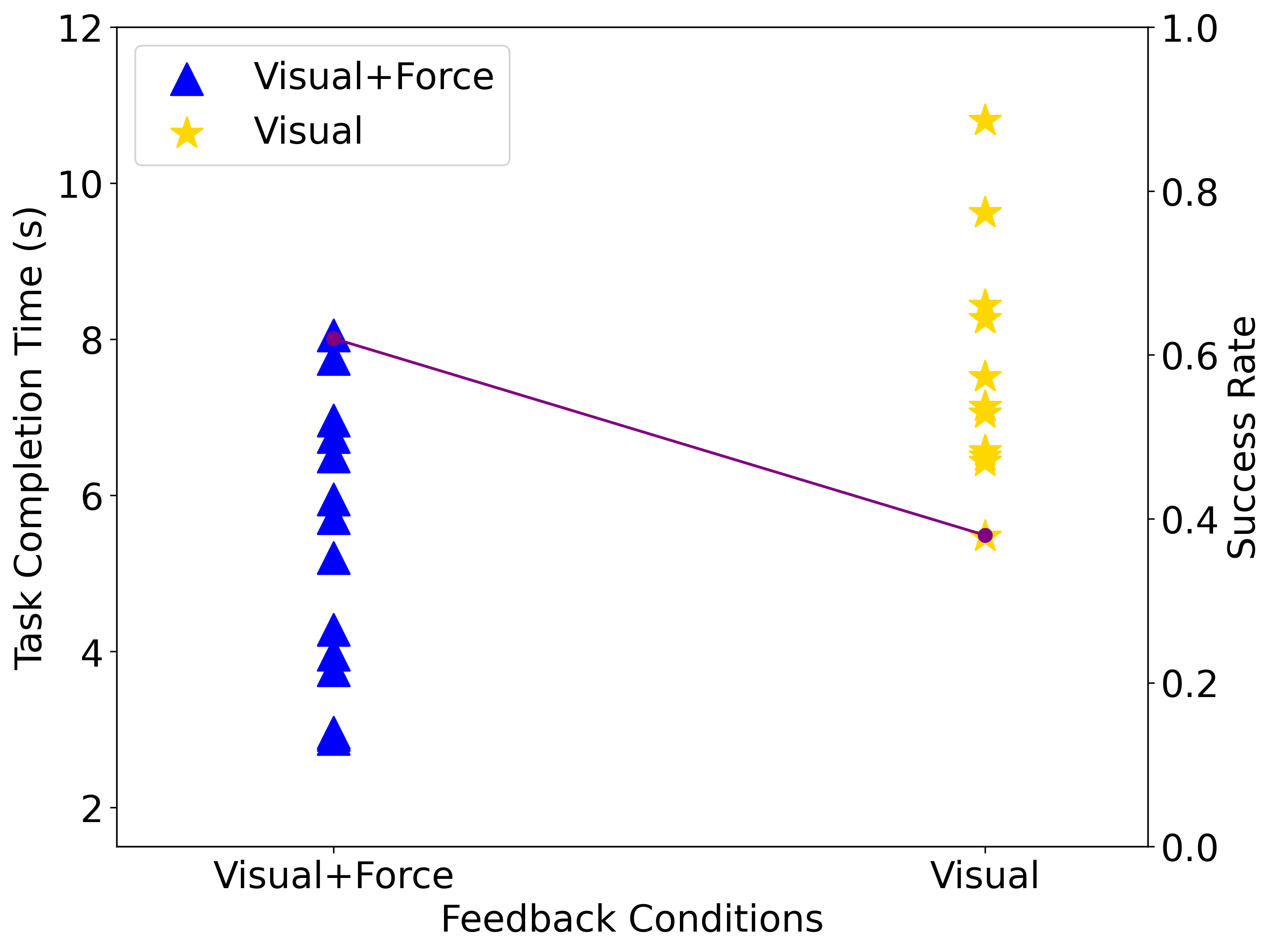}
\caption{Active sliding completion time and task success rate under two feedback conditions. The red dots represent the success rate.}
\label{fig:activeslide}
\end{figure}
In this experiment, the operator grasped a marker pen and placed it in the initial state in Fig.\ref{fig:exper} (a), then to the next intermediate state in Fig. \ref{fig:exper} (b), and finally to the last expected state in Fig. \ref{fig:exper} (c). By controlling the force applied to the pen, the operator could change the friction force between the pen and the fingertips, leading the pen to rotate at different speeds under the effect of gravity. We evaluated two conditions: visual feedback only and visual plus haptic feedback. We performed twenty-five repetitions of the operation and measured the task completion time and success rate.

The experimental results show that the architecture with visual plus haptic feedback significantly reduced the operation time and improved the success rate. The rotation speed of the pen was the decisive factor affecting the sliding speed. In the expected position, the speed was expected to zero, while in other positions, the pen was expected to rotate as fast as possible. The force applied to the pen determined its rotation speed. 

In the condition with only visual feedback, the operator had to rely on indirect reasoning  about the grip force based on visual information to adjust the force applied to the pen. The operator used their prior experience to control the force based on the observed rotation speed and position of the pen. However, this control process was often delayed and uncertain.
The haptic feedback enabled the operator to receive direct force feedback from the dexterous hand, avoiding the need for delayed and uncertain reasoning based on visual information. The operator could make accurate decisions about the force applied to the pen, leading to improved performance.
\subsubsection{Deformable Object Grasping}

To evaluate the performance of our system in deformable object grasping task, we selected a low-hardness soft plastic bottle and measured the deformation of each indentation relative to the normal state and accumulated them to evaluate the deformation during the grasping process. We adopted two strategies. One is an aggressive strategy, where the operator was instructed to complete the grasping task as quickly as possible. This strategy aims to complete the task quickly but may cause large deformation of the object. The other is a conservative strategy, Where the operator was instructed to complete the task with the minimum deformation. This strategy aims to protect the object from damage but may take longer time to complete.


\begin{figure}[tbp]  
    \centering        
    \includegraphics[width=0.48\textwidth]{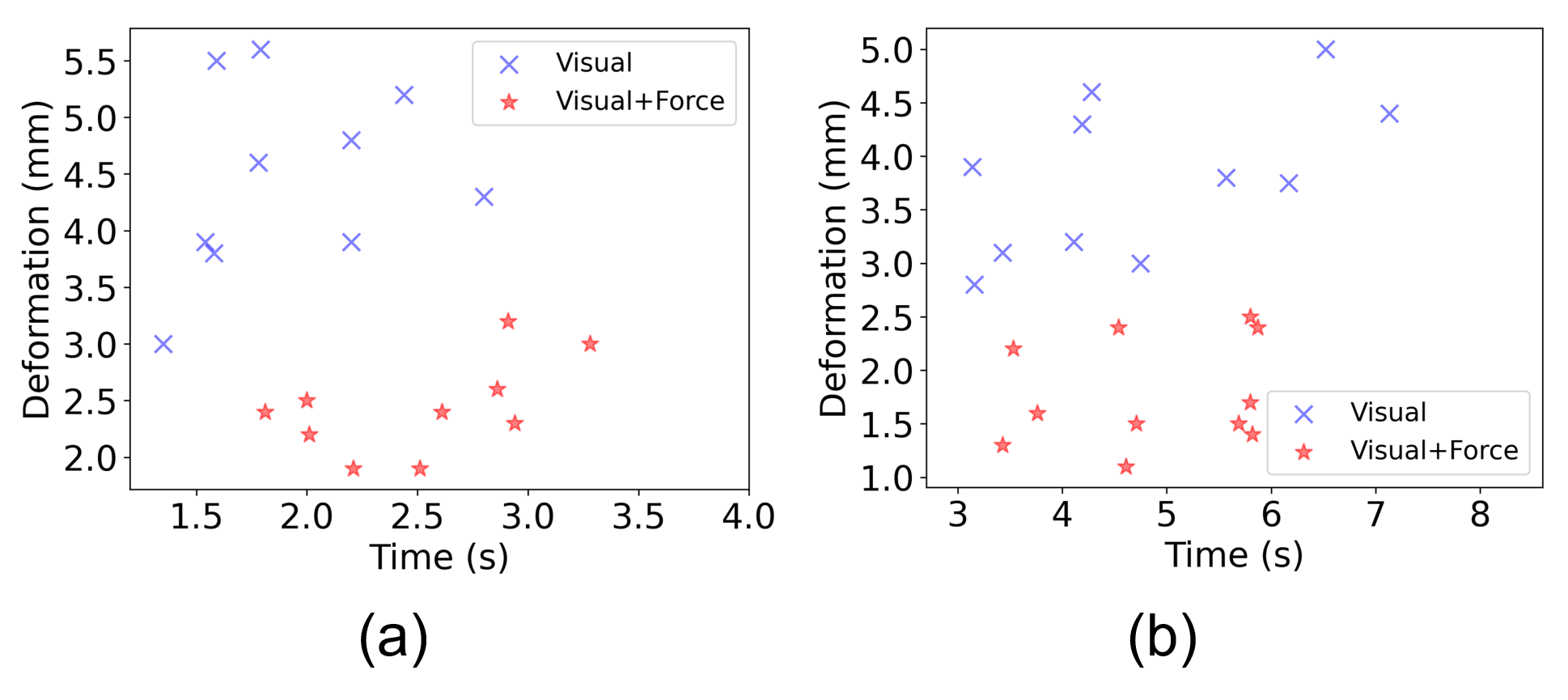}  
    \caption{Grasping deformable object completion time and deformation under two
feedback conditions with different strategy. (a) Aggressive strategy. (b) Conservative strategy.}  
    \label{fig:gr}  
\end{figure}
The experimental results in Fig. \ref{fig:gr} show that the use of visual and haptic feedback together resulted in smaller deformation values for both strategies. As shown in Table \ref{tab:grasping_results}. In the aggressive grasping strategy, the average time for grasping with visual plus haptic feedback was 2.514 s, and the average deformation was 2.44 mm. In contrast, the average time for grasping with only visual feedback was 1.96 s, and the average deformation was 4.46 mm. In the conservative grasping strategy, the average time for grasping with visual plus haptic feedback was 5.356 s, and the average deformation was 1.927 mm. In contrast, the average time for grasping with only visual feedback was 5.245 s, and the average deformation was 4.18 mm. This deformation was often small and difficult to observe visually, while with haptic feedback, the operator can feel the touch and then stop grasping. This resulted in smaller deformation for grasping with visual and haptic feedback compared to grasping with only visual feedback.

\begin{table}[h]
    \centering
    \caption{Experimental Results of Deformable Object Grasping}
    \resizebox{0.48\textwidth}{!}{ 
        \begin{tabular}{@{}lcccc@{}}
            \toprule
            Grasping Strategy & Feedback Type          & Average Time (s) & Average Deformation (mm) \\ \midrule
            Aggressive        & Visual + Haptic        & 2.514             & 2.44                     \\
                             & Visual Only            & 1.96              & 4.46                     \\ \midrule
            Conservative      & Visual + Haptic        & 5.356             & 1.927                    \\
                             & Visual Only            & 5.245             & 4.18                     \\ \bottomrule
        \end{tabular}
    }
    \label{tab:grasping_results}
\end{table}

\subsection{User Teleoperation Performance Survey}
To investigate the effect of haptic feedback on reducing the cognitive burden of operators and improving the teleoperation efficiency of bimanual robots, we conducted a user study. Ten participants were invited to participate in the study, and none of them had experience with VR, exoskeletons, or teleoperation.

Participants were required to use the telerobotic system to perform teleoperation tasks, and their completion times were recorded. Subsequently, participants were asked to complete a subjective survey to evaluate their user experience with the system's performance.

A task was selected from a daily life scenario, where the dual-arm and robotic hand was placed at the initial position, and the operator was required to control the left and right hands to complete the task of transporting fragile fruit. A plastic basket was placed on the left side, and the operator needed to lift it up to a suitable height to place the fruit. A peeled orange was placed on the right side, and the operator was required to command the hand to grasp the fruit without breaking it and move it to the basket. When the fruit was successfully transported without being broken, the operator's completion time was recorded.

\begin{figure}[tbp]  
    \centering        
    \includegraphics[width=0.48\textwidth]{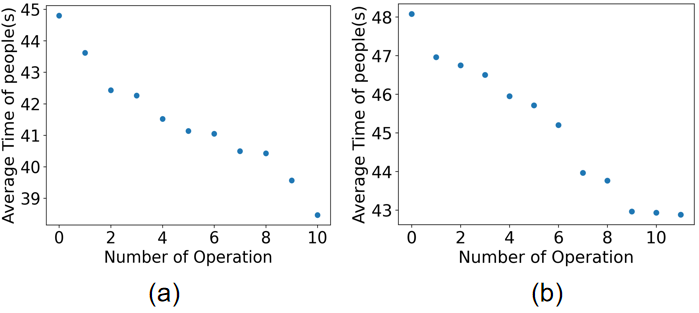}  
    \caption{Average time for ten people to complete the operation task each time under different sensory feedback conditions. (a) Visual + Haptic Feedback.(b)Visual Feedback.}  
    \label{fig:us}  
\end{figure}

Ten participants were required to perform the task under two conditions: visual feedback and visual plus haptic feedback, until they successfully completed the task ten times. 
We calculated the average completion time of ten people for each operation under each condition.
According to the statistical analysis of the participants' completion time Fig. \ref{fig:us}, the condition with visual plus haptic feedback resulted in shorter completion times, and the participants were able to adapt to the device faster and reduce the task execution time, improving the operation efficiency.
Subsequently, the operators were asked to complete a subjective survey. The questions asked were:
\begin{itemize} 
\item Question 1:Which sensory feedback method gave you a stronger sense of immersion? 
\item Question 2:Did the haptic feedback help you during the operation, reducing your cognitive burden?
\end{itemize}

{\bf{Ask and Analysis 1:}} 90\% of participants believe that the visual plus haptic feedback method provides a stronger sense of immersion and a better telepresence effect during operation. The stereo rendering of the VR glasses gives the operator a strong sense of telepresence, and the isomorphic operation method enables the operator to directly control the action through mapping. These provide the operator with an intuitive immersive experience, creating a realistic sense of telepresence in the remote scene. However, according to participants' feedback, when the operator performs grasping actions on objects, the lack of haptic feedback information makes the operator feel detached from the immersive experience. The visual plus haptic feedback method compensates for this weakness. Furthermore, the participants report that the stronger sense of immersion provided by the visual plus haptic feedback method enables them to better focus on completing the task.

{\bf{Ask and Analysis 2:}} 70\% of operators report that haptic feedback makes the process of transporting fragile fruit easier. 100\% of operators report that haptic feedback greatly helps them grasp the fruit without damaging it, especially after repeated operations. Compared to single visual feedback, the introduction of haptic feedback reduces the operators' cognitive burden of inferring grasping force and grasping quality through visual feedback. The haptic feedback assists the operators in making decisions more quickly and improving teleoperation efficiency.

\section{Conclusion and Future Work}
In this paper, we introduce haptic feedback into the bimanual telerobotic system, constructing a bilateral haptic feedback human-robot interaction interface that expands the sensory feedback modalities of teleoperation. This interface enables the operator to perceive the physical properties of objects, compensating for the limitations of visual information. While improving the efficiency of operation and expanding the range of complex tasks that can be executed by the bimanual robot, the interface also achieves more intuitive, realistic, and immersive operation, reducing the cognitive burden of the operator.

In this telerobotic system, we only provided VR visual feedback and haptic feedback information. In the future, we hope to provide more diverse feedback information, including temperature, texture, and skin stretching feedback from the remote site. By increasing the feedback information, we hope further improve the teleoperation performance of the bimanual robot, enabling the operator to execute tasks more intuitively, realistically, and immersively.

\bibliographystyle{IEEEtran} 
\bibliography{references} 
\end{document}